\documentclass[reprint, amsmath,amssymb, aps]{revtex4-2}

\usepackage{graphicx}
\usepackage{dcolumn}
\usepackage{bm}
\usepackage{hyperref}
\usepackage[utf8]{inputenc}
\usepackage[T1]{fontenc}
\usepackage{mathptmx}
\usepackage{tikz}
\usetikzlibrary{arrows,shapes}
\usetikzlibrary{decorations.markings}
\usetikzlibrary{calc,decorations.markings,arrows.meta}
\usepackage{etoolbox}

\makeatletter
\def\@email#1#2{%
 \endgroup
\patchcmd{\titleblock@produce}
  {\frontmatter@RRAPformat}
 {\frontmatter@RRAPformat{\produce@RRAP{*#1\href{mailto:#2}{#2}}}\frontmatter@RRAPformat}
  {}{}
}
\makeatother

\begin{document}

\preprint{}

\title{Osculatory Dynamics: Framework for the Analysis of Oscillatory Systems}

\author{Marco Thiel}
 \email{m.thiel@abdn.ac.uk}
\affiliation{%
Department of Physics\\
University of Aberdeen (UK)
}%

\date{\today}

\begin{abstract}

Intractable phase dynamics often challenge our understanding of complex oscillatory systems, hindering the exploration of synchronisation, chaos, and emergent phenomena across diverse fields. We introduce a novel conceptual framework for phase analysis, using the osculating circle to construct a co-moving coordinate system, which allows us to define a unique phase of the system. This coordinate independent, geometrical technique allows dissecting intricate local phase dynamics, even in regimes where traditional methods fail. Our methodology enables the analysis of a wider range of complex systems which were previously deemed intractable.

\end{abstract}

\maketitle

\section{\label{sec:level1}Introduction}

The study of oscillatory systems encompasses a diverse range of behaviours, from periodic to quasi-periodic, chaotic and stochastic systems. These systems often present complex trajectories in phase space, requiring sophisticated techniques for analysing their instantaneous states, e.g. their instantaneous phase. Traditional approaches such as using Hilbert Transforms to determine the phase, while effective in certain contexts, can struggle with the complex nature of these systems. In this work, we introduce a novel, geometrically inspired methodology using the osculating circle, a fundamental concept in differential geometry. This technique, which involves constructing a co-moving coordinate system based on the osculating circle, allows for a precise extraction of instantaneous phase and frequency information across various types of oscillatory systems. Our approach overcomes some limitations of conventional methods, offering a robust, coordinate-independent framework for analysing intricate dynamics, as evidenced in our exploration of chaotic dynamics \cite{Strogatz1994, Alligood1996} and other complex behaviours \cite{Pikovsky2001, Rosenblum1997}, thereby extending our ability to study a new range dynamical systems.\\
The subsequent sections of this paper introduce the theoretical underpinnings of our methodology, and illustrate its application in a variety of oscillatory systems. We begin with a detailed exposition of the mathematical principles behind the osculating circle approach and how to define a phase and frequency based on it, followed by a series of case studies demonstrating its effectiveness. Comparative analyses with existing methods to define the phase are presented to highlight the enhanced capabilities of our approach. The paper concludes with a discussion on the broader implications of our findings and potential future directions in the study of dynamical systems.

\section{\label{sec:level2}Theoretical Background}

Traditional phase analysis methods often face constraints when applied to complex oscillatory systems: the definition of the phase itself can pose the first challenge. One common approach involves projecting the system onto a plane and defining a central point, around which a rotational orbit is established. The phase of this orbit can then be determined in polar coordinates. Another frequently used approach uses the Hilbert transform on an arbitrarily chosen component of the oscillatory system. While these techniques are effective in certain contexts, they encounter difficulties with non-phase-coherent systems and certain chaotic oscillators \cite{romano2020detection,kralemann2011reconstructing,rosenblum2001detecting}. Challenges include assumptions about the system's dynamics, reliance on phase reduction theory primarily suited for weakly coupled limit cycle oscillators, and finding approaches to accurately computing a phase for systems lacking a clear rotational centre. Additionally, traditional methods often fall short in handling non-stationary data, leading to an incomplete understanding of synchronisation phenomena in complex scenarios.
Our approach circumvents these limitations by utilising the geometrical properties of osculating circles. This technique offers a robust, coordinate-independent method for phase and frequency analysis, particularly effective in systems that are non-phase-coherent and non-stationary. Unlike traditional methods, our framework does not require predefined assumptions, heuristics or case by case projections, providing a versatile tool for elucidating the complex behaviour of oscillatory systems.
The core of our methodology lies in conceptualising the instantaneous phase of a trajectory in a dynamical system. This is achieved by visualising it as the angle parameter of the circular motion that locally best approximates the system's path at any given point. The osculating circle, which mathematically 'kisses' the trajectory at the point of tangency, sharing the same first and second derivatives, becomes the cornerstone of this analysis.

\subsection{Osculating Circles in 3 Dimensions}

In this section, we introduce the mathematical foundations of the osculating circle, particularly focusing on the three-dimensional case. Central to this discussion is the concept of the "moving trihedral," also known as the "accompanying Dreibein", which is a fundamental concept in differential geometry.

\begin{figure}[ht]
    \centering
    \includegraphics[width=0.4\textwidth]{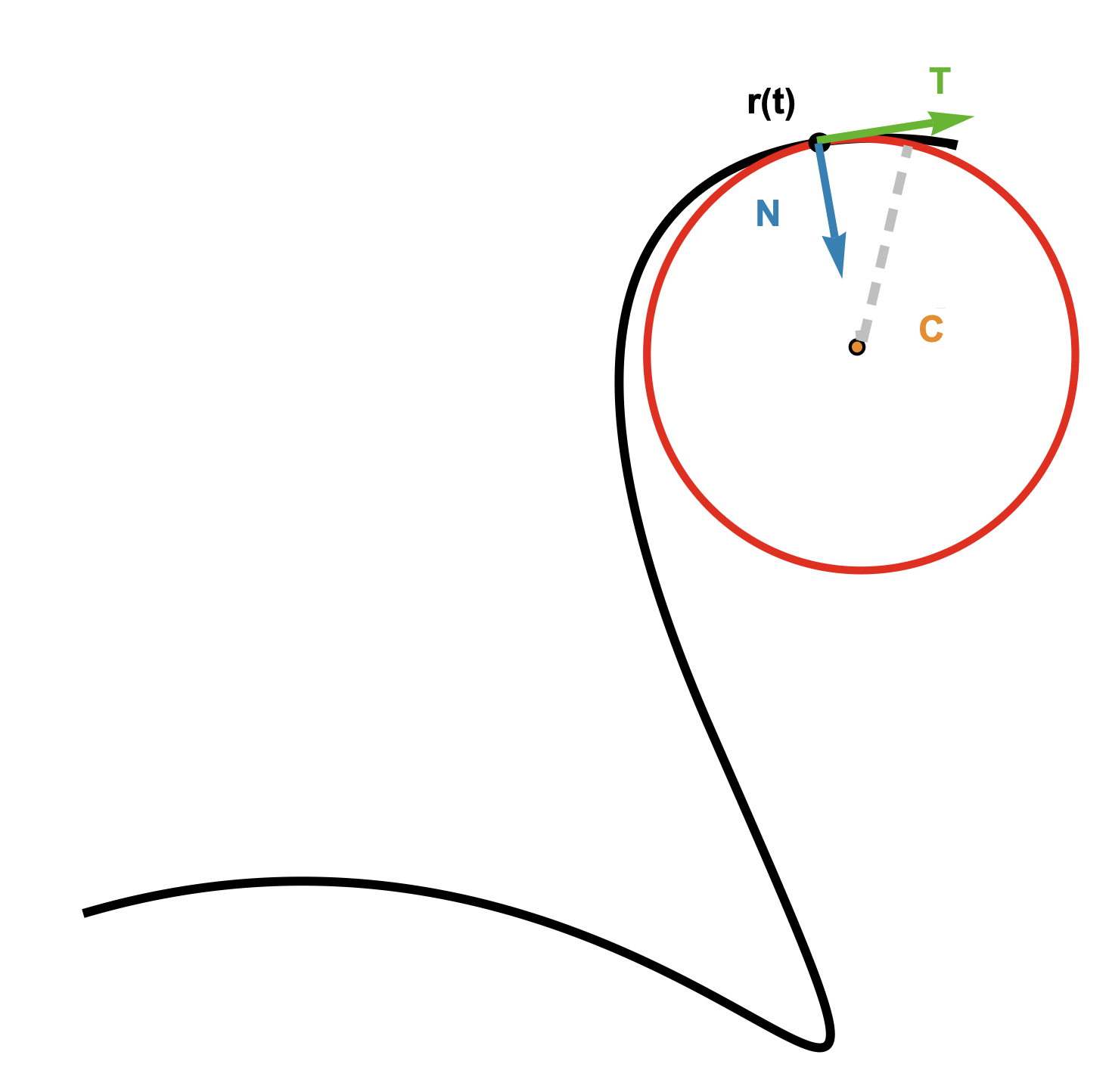}
 \caption{Illustration of the osculating circle at a point $\mathbf{r}(t)$ on a trajectory with the unit tangent vector $\mathbf{T}$ and the unit normal vector $\mathbf{N}$ . The radius of the osculating circle is indicated by the dashed grey line. $\mathbf{C}$ is the centre of the osculating circle.}
\label{fig:osculating}
\end{figure}

Consider a smooth curve \( \mathbf{r}(t) \) parameterized by time \( t \). To facilitate our analysis, we introduce the arc length parameter \( s \), defined as:

\begin{equation}\label{def:arclength}
s(t) = \int_{t_0}^{t} \|\mathbf{\dot{r}}(\tau)\| \, d\tau
\end{equation}

where \( \mathbf{\dot{r}}(t) \) denotes the derivative of \( \mathbf{r}(t) \) with respect to time \( t \). The arc length \( s \) provides a natural parameterization of the curve, invariant under reparametrization.

The moving trihedral consists of three mutually orthogonal unit vectors: the unit tangent vector \( \mathbf{T}(s) \), the unit normal vector \( \mathbf{N}(s) \), and the binormal vector \( \mathbf{B}(s) \). These vectors are defined as follows:

\begin{itemize}
    \item The unit tangent vector \( \mathbf{T}(s) \) is given by:
    \[
    \mathbf{T}(s) = \frac{d\mathbf{r}}{ds}
    \]
    \item The unit normal vector \( \mathbf{N}(s) \) is defined as:
    \[
    \mathbf{N}(s) = \frac{d\mathbf{T}}{ds} \bigg/ \left\| \frac{d\mathbf{T}}{ds} \right\|
    \]
    \item The binormal vector \( \mathbf{B}(s) \) is given by the cross product of \( \mathbf{T}(s) \) and \( \mathbf{N}(s) \):
    \[
    \mathbf{B}(s) = \mathbf{T}(s) \times \mathbf{N}(s)
    \]
\end{itemize}

These vectors satisfy Frenet's Formulas, which describe how the trihedral rotates as we move along the curve. The formulas are:

\[
\frac{d\mathbf{T}}{ds} = \kappa(s) \mathbf{N}(s)
\]
\[
\frac{d\mathbf{N}}{ds} = -\kappa(s) \mathbf{T}(s) + \tau(s) \mathbf{B}(s)
\]
\[
\frac{d\mathbf{B}}{ds} = -\tau(s) \mathbf{N}(s)
\]

Here, \( \kappa(s) \) is the curvature, and \( \tau(s) \) is the torsion of the curve. The curvature \( \kappa(s) \) is a measure of how sharply the curve bends and is defined as:

\[
\kappa(s) = \left\| \frac{d\mathbf{T}}{ds} \right\|
\]

To relate derivatives with respect to the arc length $s$ to derivatives with respect to time  $t$, we can use Eq.(\ref{def:arclength}) and the chain rule from calculus:

\[
ds = \|\mathbf{\dot{r}}(t)\| dt.
\]

Therefore, the relationship between the derivatives is:
\[
\frac{d}{ds} = \frac{1}{\|\mathbf{\dot{r}}(t)\|} \frac{d}{dt}.
\]

Applying this relationship we obtain expressions as functions of the time $t$ for the unit tangent vector:

\[
\mathbf{T}(t) = \frac{\mathbf{\dot{r}}(t)}{\|\mathbf{\dot{r}}(t)\|},
\]

the normal vector $\mathbf{N}$

\[
\mathbf{N}(t) = \frac{\|\dot{\mathbf{r}}(t)\|^2 \ddot{\mathbf{r}}(t) - (\dot{\mathbf{r}}(t) \cdot \ddot{\mathbf{r}}(t)) \dot{\mathbf{r}}(t)}{\left\| \|\dot{\mathbf{r}}(t)\|^2 \ddot{\mathbf{r}}(t) - (\dot{\mathbf{r}}(t) \cdot \ddot{\mathbf{r}}(t)) \dot{\mathbf{r}}(t) \right\|},
\]

and the curvature $\kappa(s)$ in terms of $t$:

\begin{equation}\label{def:kappa}
\kappa(t) = \frac{\|\mathbf{\dot{r}}(t) \times \mathbf{\ddot{r}}(t)\|}{\|\mathbf{\dot{r}}(t)\|^3}.
\end{equation}

The osculating circle at a point \( \mathbf{r}(t) \) on the curve is the circle that best approximates the curve at that point. Its center \( \mathbf{C}(t) \) is located in the direction of the normal vector \( \mathbf{N}(t) \) at a distance equal to the radius of curvature \( R(t) = \frac{1}{\kappa(t)} \) from the point \( \mathbf{r}(t) \):

\[
\mathbf{C}(t) = \mathbf{r}(t) + \frac{1}{\kappa(t)} \mathbf{N}(t).
\]

This construction provides a geometrical framework for analysing the local behaviour of the curve. The osculating circle "kisses" the curve at \( \mathbf{r}(t) \), sharing the same first and second derivatives, and thus captures the local curvature properties of the trajectory.

For the special case of circular motion with constant angular speed $\omega$, the following equation holds $\omega_{\text{circ}}=\frac{\|\mathbf{\dot{r}}(t)\|}{R}$. 
We propose to use the analogous definition of angular speed for the general case of any smooth trajectory in phase space:

\begin{equation}\label{def:omega}
\omega(t) = \kappa(t) \|\mathbf{\dot{r}}(t)\|.
\end{equation}

Note that in the general case, $\omega(t)$ will be a function of time. 

We then use this Eq.(\ref{def:omega}) to define an instantaneous phase based on the construction of the osculating circle by integrating \(\omega(t)\) over time:
\begin{equation}\label{def:phase}
    \phi(t) = \int_{t_0}^{t} \omega(\tau) d\tau,
\end{equation}
where \(t_0\) is the initial time.

Using the osculating circle to define $\omega$ and then the phase $\phi$ is the central idea to our method, enabling a precise and geometrical approach to phase analysis in oscillatory systems.

\subsection{Osculating Circles in higher Dimensions \label{sec:highdim}}

Extending our methodology to higher-dimensional manifolds, we consider a trajectory \(\mathbf{r}(t)\) within an \(n\)-dimensional manifold. Here, \(\mathbf{r}(t)\) represents a vector in \(\mathbb{R}^n\), with \(t\) as time. The tangent vector \(\mathbf{v}(t) = \frac{d\mathbf{r}}{dt}\) encapsulates the local motion along the curve. In differential geometry, these entities can be expressed in tensorial form, a necessary abstraction for higher dimensions.

The curvature tensor, \(\kappa_{ij}(t)\), is central to understanding trajectory bending in the manifold. It is defined by the covariant derivative:
\begin{equation}
    \kappa_{ij}(t) = \nabla_i v_j - \nabla_j v_i
\end{equation}
with \(\nabla_i\) representing the covariant derivative concerning coordinate \(x^i\) 
\begin{equation}
\nabla_i v_j = \partial_i v_j + \Gamma^k_{ij} v_k,
\end{equation} 
and where 
\begin{equation}
\Gamma^{k}_{ij} = \frac{1}{2} g^{kl} \left( \partial_j g_{il} + \partial_i g_{jl} - \partial_l g_{ij} \right)
\end{equation}
are the Christoffel symbols of the second kind. As usual repeated indices imply summation (Einstein summation convention). The tensor $\kappa_{ij}(t)$ differs from the simpler curvature in lower dimensions, e.g. it is not a scalar, reflecting the complexity of trajectory bending in higher-dimensional spaces.

The norm of this curvature tensor is computed using the metric tensor \(g_{ij}\) of the manifold:
\begin{equation}
    \|\kappa(t)\| = \sqrt{g^{ik}g^{jl}\kappa_{ij}(t)\kappa_{kl}(t)}
\end{equation}
Here, \(g^{ik}\) is the inverse of \(g_{ik}\). This computation is more demanding than its lower-dimensional counterpart due to the involvement of the metric tensor and the tensorial nature of curvature.

The speed of the tangent vector \(\mathbf{v}(t)\) is given by:
\begin{equation}
    \|\mathbf{v}(t)\| = \sqrt{g_{ij}v^i(t)v^j(t)}
\end{equation}
involving the metric tensor.

The angular speed \(\omega(t)\) remains a pivotal measure, representing the 'rotation speed' of the trajectory within the manifold:
\begin{equation}\label{eq:omegaHD}
    \omega(t) = \|\kappa(t)\|  \cdot \|\mathbf{v}(t)\|
\end{equation}
Integrating this angular speed over time, as before in Eq.(\ref{def:phase}), we obtain the phase \(\phi(t)\) in the \(n\)-dimensional case.
This tensorial approach provides a general tool for phase analysis in higher-dimensional systems, where other methods cannot be easily applied. We illustrate the application this approach in section \ref{sec:higherDims}.

\section{\label{sec:level3}Methodology}

Our methodology for determining the osculating circle and the associated instantaneous phase involves several key steps, adapted to handle the complexities of both low and high-dimensional systems. This section details the process for 3-dimensional systems and $n$-dimensional systems, where the system is defined by a set of ODEs, as well as for time series data.
If the system is defined by differential equations, we proceed with the following steps. First, we numerically integrate the differential equations governing the dynamical system to obtain the trajectory,  $r(t)$. Once the trajectory is obtained, we compute the angular speed, $\omega(t)$, and the instantaneous phase, $\phi(t)$, using Eqs.(\ref{def:omega})-(\ref{def:phase}) for three dimensional systems, and Eq.(\ref{eq:omegaHD}) and Eq.(\ref{def:phase}) for the $n$-dimensional case. This involves calculating the curvature $\kappa(t)$ and the radius $R(t)$ of the osculating circle.

For systems provided as a time series (i.e., discrete points) we first interpolate between the discrete data points to form a continuous curve, which is crucial for accurately computing the derivatives needed for $\kappa(t)$ and $\omega(t)$. If the time series is compromised by noise, denoising techniques might also need to be applied to mitigate the effects of noise in the data. Based on the interpolated and de-noised curve, we can use embedding techniques \cite{Ott2002} and then compute the curvature $\kappa(t)$, angular speed $\omega(t)$, and instantaneous phase $\phi(t)$ as previously described.

\section{Application to The R\"ossler System}

In this section, we apply our osculating circle methodology to the R\"ossler system \cite{Rossler1976}, a classic model in chaotic dynamics. The R\"ossler system, described by the following differential equations,

\begin{align}
    \dot{x} &= -y - z, \notag \\
    \dot{y} &= x + ay, \label{roesslereqns} \\
    \dot{z} &= b + z(x - c)\notag,
\end{align}

provides a test case for demonstrating the efficacy of our approach. Here, $a$, $b$, and $c$ are system parameters.

\begin{figure}[ht]
    \centering
    \includegraphics[width=0.3\textwidth]{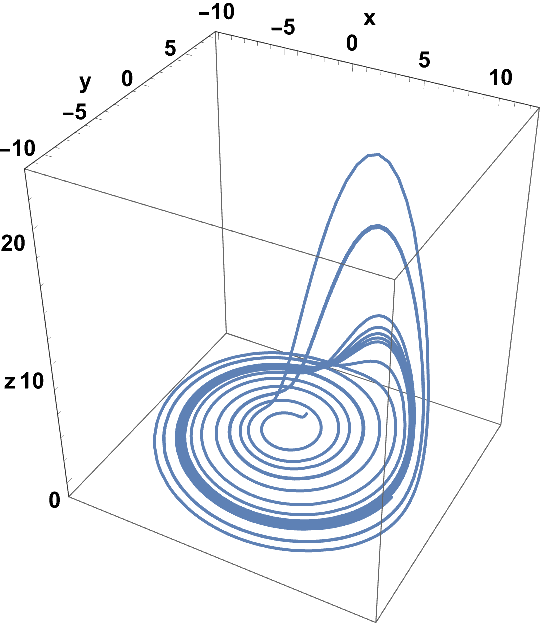}
    \caption{Trajectory of the R\"ossler attractor with parameters $a = 0.2, b = 0.2, c = 5.7$.}
    \label{fig:rossler-trajectory}
\end{figure}

We first compute the trajectory of the R\"ossler system by numerically integrating these differential equations. From the trajectory data, we then compute the angular speed $\omega(t)$ (Eq.(\ref{def:omega})), and the instantaneous phase $\phi(t)$ (Eq.(\ref{def:phase}).

Figure~\ref{fig:rossler-trajectory} displays the trajectory of the R\"ossler attractor with parameters $a = 0.2$, $b = 0.2$, and $c = 5.7$. Traditionally, in order to compute the phase, one can project the system onto the $xy$-plane, and it will oscillate about the origin. However, instead of relying on a phase definition based on this projection, we utilise the osculating circle method.

\begin{figure*}[ht]
    \centering
    \includegraphics[width=\textwidth]{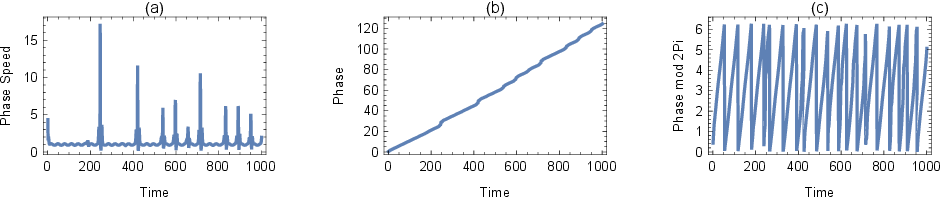}
    \caption{Detailed analysis of the R\"ossler system Fig\ref{fig:rossler-trajectory}. (a) Instantaneous phase speed of the R\"ossler system. (b) Integral of the speed; instantaneous phase. (c) Instantaneous phase modulo $2\pi$.}
    \label{fig:detailed-analysis}
\end{figure*}

Figure~\ref{fig:detailed-analysis} shows the results of the phase analysis for the R\"ossler system. Panel (a) presents the instantaneous angular speed $\omega(t)$, while panel (b) shows the integrated phase $\phi(t)$. Importantly, the phase $\phi(t)$ is a monotonically increasing scalar function, which is a critical feature for the definition of a phase as non-monotonous functions could lead to ambiguities \cite{Pikovsky2001}. Panel (c) displays the phase modulo $2\pi$.

A key observation is that using the right-hand sides of the differential equations, $\omega$ can be expressed as a function of the coordinates alone, see Eqs.(\ref{kappaofcoords}-\ref{omegaofcoords}) . This means we can compute the angular speed for each point in the phase space using only the coordinates, without needing to solve the ODEs. 

\begin{figure*}[ht]
    \centering
    \includegraphics[width=0.7\textwidth]{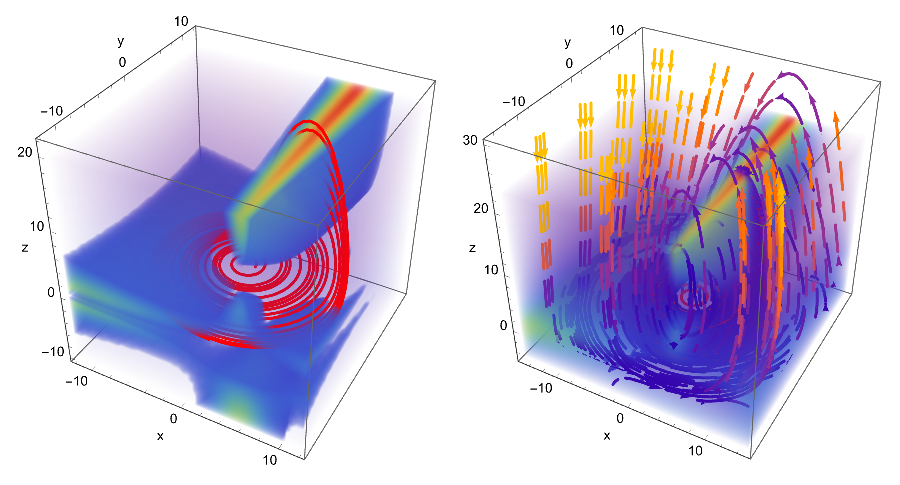}
    \caption{Angular speed representation as a function of phase space coordinates for the R\"ossler system.  (Left) Illustration of the speed, with cooler colours and higher transperancy indicating lower angular speeds and warmer colours indicating higher speeds. (Right)  same as (left) but also shows the flow dynamics using streamlines.}
    \label{fig:phasespacespeed}
\end{figure*}

Figure \ref{fig:phasespacespeed} visually demonstrates the angular speed in the Rössler system's phase space. In this figure, warmer colours indicate higher angular speeds ($\omega$), while cooler colours and the more transparent regions indicate lower angular speeds; for reference the attractor is shown by a red trajectory. The left panel of Figure \ref{fig:phasespacespeed} illustrates the speed distribution across the phase space, with regions of higher speed (depicted in red) aligning with areas of significant curvature and points where the trajectory diverges from the xy-plane. This shows the instantaneous angular speed for motion on the attractor.

Figure \ref{fig:phasespacespeed} is particularly interesting because it helps us understand the system's behaviour when perturbed away from the attractor. The right panel of Figure \ref{fig:phasespacespeed} shows the flow using streamlines. When the system is perturbed upwards from the rotational motion in the xy-plane, it enters a transparent region, indicating very low angular speed. As expected, the system moves directly back to the xy-plane with minimal rotation.

These images can enhance our understanding of the process of synchronisation. For instance, consider two uncoupled R\"ossler systems following their attractors. When these systems interact at a certain coupling strength, they adjust their phases and angular speeds. The interaction effectively moves the systems into areas where their angular speeds are very similar or the same. This adjustment is illustrated in Figure \ref{fig:phasestotalvspart}. The systems align their bursts of angular speed, leading to synchronised motion.

\begin{figure*}[ht]
    \centering
    \begin{minipage}{0.43\linewidth}
        \includegraphics[width=\linewidth]{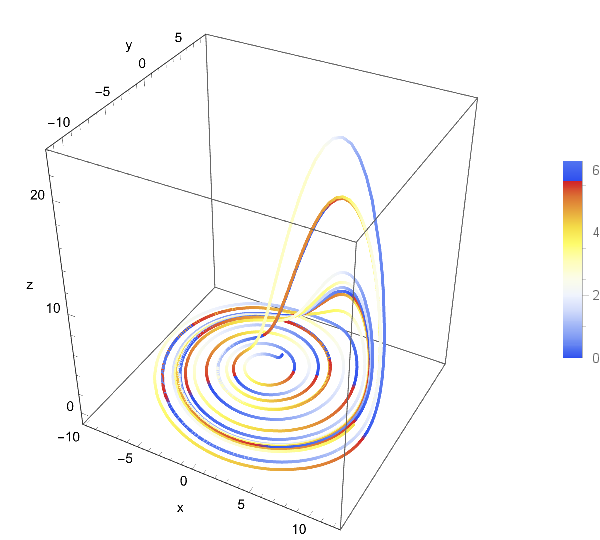}
        \caption{Attractor coloured by instantaneous phase. There is no obvious pattern, e.g., relationship to the angle in the $xy$ plane.}
        \label{fig:instant-phase}
    \end{minipage}
    \hfill
    \begin{minipage}{0.53\linewidth}
        \includegraphics[width=\linewidth]{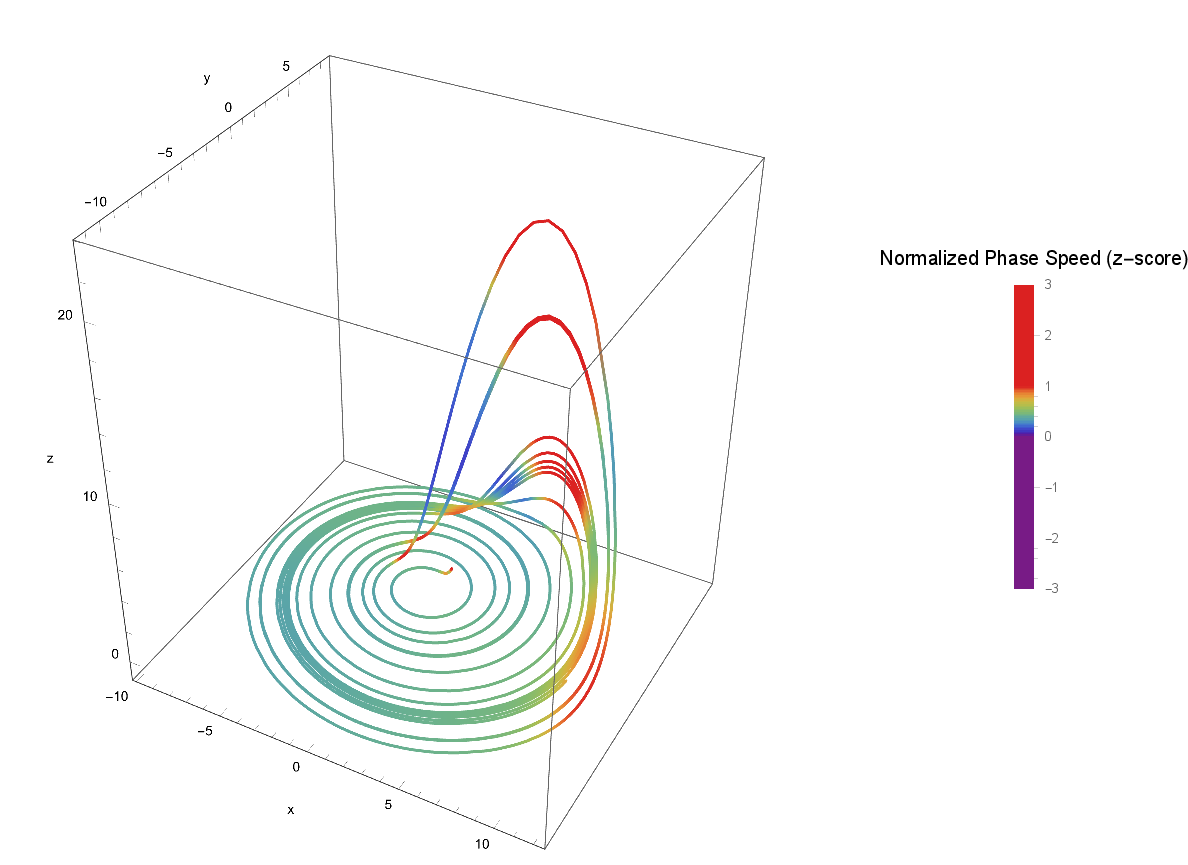}
        \caption{Attractor coloured by angular speed. The colour scheme has been appropriately rescaled.}
        \label{fig:phase-angular-speed}
    \end{minipage}
\end{figure*}

Figure~\ref{fig:instant-phase} presents the R\"ossler attractor where each point of the trajectory is colour coded according to its instantaneous phase. This leads to an irregular colour pattern, highlighting the curvature-dependent nature of our phase calculation. This is a notable departure from traditional phase methods like the Hilbert Transform or methods based on the projection on a 2-dimensional plane, where uniform colouring often corresponds to similar directions in that plane.

Figure~\ref{fig:phase-angular-speed} illustrates the attractor coloured by angular speed, and in contrast to Fig.~\ref{fig:instant-phase} displays remarkable coherence. This coherence reflects our method's sensitivity to changes in curvature, making it necessary to adjust the colour scale to clearly delineate these variations. Notably, areas of the trajectory with high curvature, or equivalently, smaller radii of the osculating circle, exhibit higher angular speeds.

This comparative analysis brings to light the differences and some of the advantages of our curvature-based phase computation. It helps to reveal detailed dynamical aspects of the system and links it closely to the geometry of the attractor, which might remain obscured with conventional phase calculation methods.

\begin{figure*}[ht]
    \centering
    \includegraphics[width=0.8\textwidth]{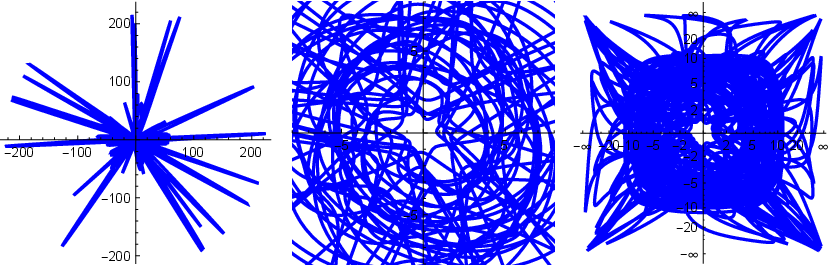}
    \caption{Representation of the phase dynamics of the R\"ossler attractor in the reference frame of the osculating circle. (left) System trajectory in the reference frame of the osculating circle. Note the large excursions of the trajectory, which are a result of areas with high curvature. (middle) Representation of the region close to the origin of the osculating circle. The trajectory moves around the centre, indicating the existence of a centre for the oscillation. (right) Same as (left) but with axes transformed to display the interval from $(-\infty, \infty)$. This representation shows that there is a rotation about the origin, which allows us to define a phase $\phi(t)$. The position $x(t) = r(t) \exp(i \phi(t))$ allows for the definition of a monotonously increasing phase $\phi(t)$, which is essential for phase synchronisation analysis. The axes represent coordinates in the coordinate system of the osculating circle. Different curves indicate various trajectory components.}
    \label{fig:phaseinoscullatingreferenceframe}
\end{figure*}

Visualising the trajectory's motion in the reference frame of the osculating circle offers an intuitive representation of oscillation. By leveraging the radius of the osculating circle along with the phase, we can effectively visualise the dynamics of the phase in Figure~\ref{fig:phaseinoscullatingreferenceframe}. The trajectory shown in the figure is given by the real and imaginary parts of  $x(t) = r(t) \exp(i \phi(t))$, where  $r(t)$ is the radius of the osculating circle and $\phi(t)$ is the phase computed via the osculating circle.  The two panels in Figure~\ref{fig:phaseinoscullatingreferenceframe} represent the entire phase dynamics within the osculating circle's reference frame. Panel (a) displays the overall phase dynamics, while panel (b) focuses on the region around the osculating circle's origin. Here, the trajectory's large excursions, particularly evident in regions of high curvature, are consistent with the R\"ossler system's behaviour. These excursions correlate with the red areas in Figure~\ref{fig:phase-angular-speed}, indicating high angular speeds. 

Importantly, our approach allows for the calculation of angular speed, $\omega(t)$, directly from the system's coordinates, a significant advancement over traditional methods. By using the known differential equations of the R\"ossler system (Eqns \ref{roesslereqns}), we can derive the expressions for the velocity vector $\mathbf{v}(t)$ and the acceleration vector $\mathbf{a}(t)$. The curvature, $\kappa(t)$, is then given as a function of $x$, $y$, and $z$:

\begin{equation}\label{kappaofcoords}
\kappa(t) = \frac{\|\mathbf{v}(x(t),y(t),z(t)) \times \mathbf{a}(x(t),y(t),z(t))\|}{\|\mathbf{v}(x(t),y(t),z(t))\|^3},
\end{equation}

Subsequently, $\omega(t)$ is derived as:

\begin{equation}\label{omegaofcoords}
\omega(t) = \kappa(x(t),y(t),z(t)) \cdot \|\mathbf{v}(x(t),y(t),z(t))\|,
\end{equation}

resulting in an expression for $\omega(\cdot)$ as a function of the coordinates, which is used in Figure \ref{fig:phasespacespeed}. These expressions, though complex, are straightforward to derive and crucially enable us to visualise the angular speed across the entire phase space. This represents a substantial advantage in analysing chaotic systems, as it bypasses the need for system integration while providing a comprehensive view of the dynamics.

\section{Comparative Analysis with Traditional Techniques}

In this section, we compare three methods for determining the phase of the R\"ossler system with parameters $a = 0.2$, $b = 0.2$, and $c = 5.7$.

\subsection{Hilbert Transform Method}
The Hilbert transform is used to derive an analytic signal from a real-valued signal, which provides a way to compute the instantaneous phase. Given a real-valued signal $x(t)$, the Hilbert transform $\mathcal{H}\{x(t)\}$ is defined as:
\[
\mathcal{H}\{x(t)\} = \frac{1}{\pi} \text{P.V.} \int_{-\infty}^{\infty} \frac{x(\tau)}{t - \tau} \, d\tau
\]
where P.V. denotes the Cauchy principal value. Using the Hilbert transform, we create a complex signal $z(t) = x(t) + i \mathcal{H}\{x(t)\}$. This complex signal can then be projected onto a 2 dimensional plane, where the trajectory revolves around the origin, allowing us to define a phase via $\arctan2(\operatorname{Im}(z(t)), \operatorname{Re}(z(t)))$.
The \(\arctan2\) function is an extension of the trigonometric arctangent function, defined as \(\arctan2(y, x)\). It returns the angle \(\theta\) between the positive x-axis of a plane and the point given by the coordinates \((x, y)\) on it. This function considers the sign of both arguments to determine the quadrant of the angle, with the range of \(\theta\) being \(-\pi\) to \(\pi\), thus providing a unique angle for every point \((x, y)\), except the origin. The formal definition in piecewise form is:

\begin{equation}
\arctan2(y, x) = 
\begin{cases} 
\arctan\left(\frac{y}{x}\right) & \text{if } x > 0,\\
\arctan\left(\frac{y}{x}\right) + \pi & \text{if } x < 0 \text{ and } y \geq 0,\\
\arctan\left(\frac{y}{x}\right) - \pi & \text{if } x < 0 \text{ and } y < 0,\\
+\frac{\pi}{2} & \text{if } x = 0 \text{ and } y > 0,\\
-\frac{\pi}{2} & \text{if } x = 0 \text{ and } y < 0,\\
\text{undefined} & \text{if } x = 0 \text{ and } y = 0.
\end{cases}
\end{equation}

The \(\arctan2\) function is particularly useful for determining the phase of an oscillation because it provides a continuous range of angles, necessary for accurately characterising the phase evolution in systems with rotational symmetry.

\subsection{Projection onto a Plane Method}
In the projection method, we directly project the 3-dimensional R\"ossler system onto the $xy$ plane. The projected trajectory also revolves around the origin, which allows us to define the phase using $\arctan2(y(t), x(t))$.

\subsection{Osculating Circle Method}
In the osculating circle method, we compute the phase based on the curvature and angular speed derived from the trajectory of the R\"ossler system. The instantaneous phase is obtained by integrating the angular speed $\omega(t)$. Unlike the previous methods, this approach leverages the local geometrical properties of the trajectory, providing a detailed view of the phase dynamics.\\

\begin{figure*}[ht]
    \centering
    \includegraphics[width=\textwidth]{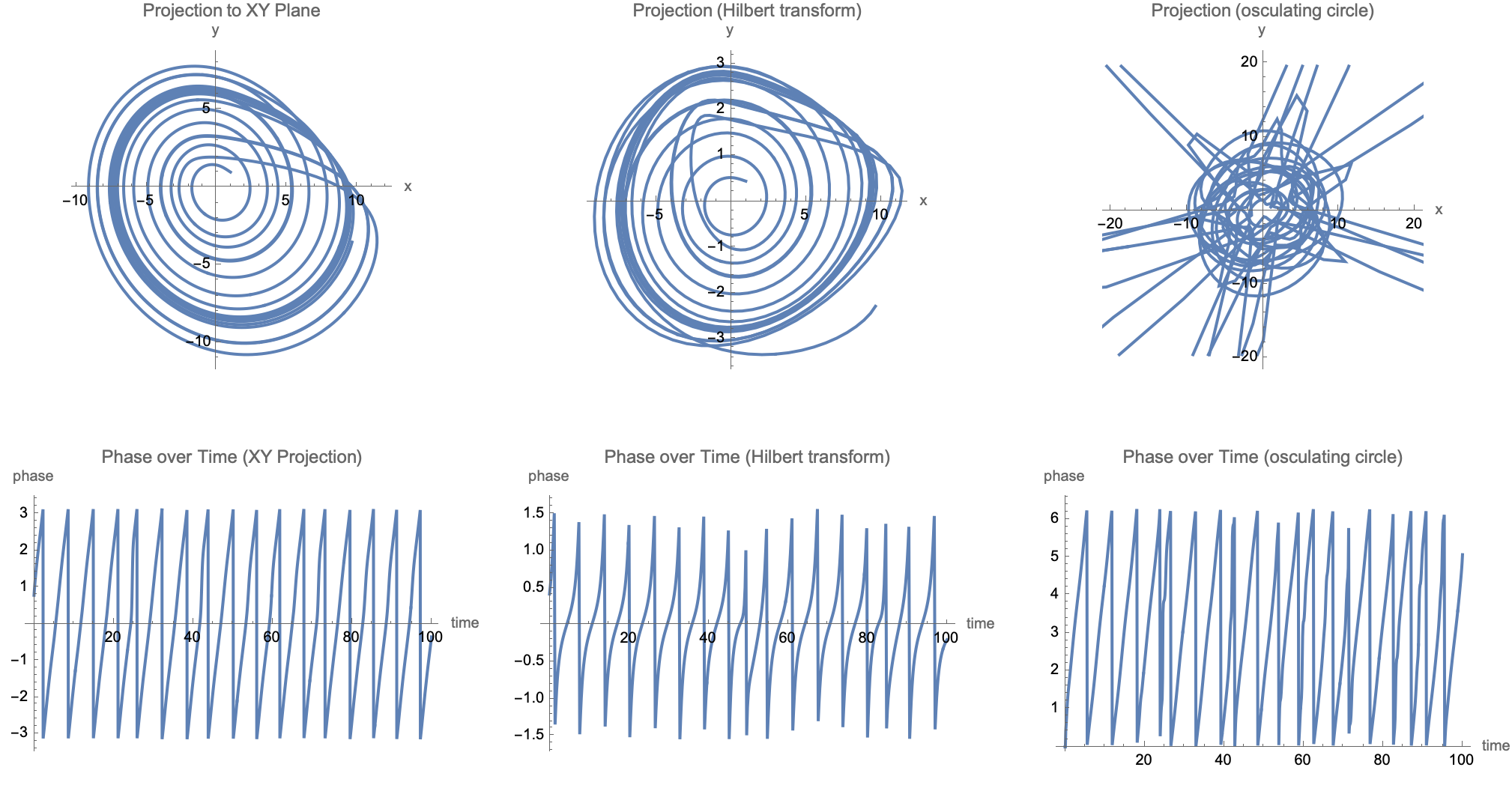}
    \caption{Comparison of different methods to determine the phase of the R\"ossler system. Top row: Planar trajectories reconstructed using (left) projection to the $xy$ plane, (middle) Hilbert transform of the $x$ component, and (right) osculating circle method. Bottom row: Corresponding phases over time using $\arctan2(y(t), x(t))$ for the first two methods and integration of $\omega(t)$ for the osculating circle method.}
    \label{fig:phasesbydefinition}
\end{figure*}

Figure \ref{fig:phasesbydefinition} shows the results of the two conventional methods to calculate the phase, alongside the results obtained with the osculating circle method. The first row presents the reconstructed planar trajectories. The leftmost panel shows the trajectory projected onto the $xy$ plane, the middle panel shows the trajectory reconstructed using the Hilbert transform of the $x$ component, and the rightmost panel shows the trajectory based on the osculating circle method.
The second row presents the corresponding phases over time. In the first two cases, the phase is defined via $\arctan2(y(t), x(t))$. In the case of the osculating circle method, the phase is obtained by integrating the angular speed $\omega(t)$. It is evident that in each case, the 2-dimensional trajectory revolves around the origin and leads to a monotonously increasing phase. However, the radii of the trajectory for the osculating circle method fluctuate much more, and likewise, the phases vary more than in the other two cases. The rapid excursions, due to changes in the osculating circle's radius appear to make the "projection" via the osculating circle less clean, but the phase definition yields a clear signal. 
This comparative analysis highlights the distinct characteristics of each method. The projection and Hilbert transform methods provide relatively smooth trajectories and phases, while the osculating circle method offers a new geometric perspective, capturing more intricate variations in the phase dynamics by emphasising the local curvature. However, in this case all three methods produce valid phase signals. 

\begin{figure*}[ht]
    \centering
    \includegraphics[width=\textwidth]{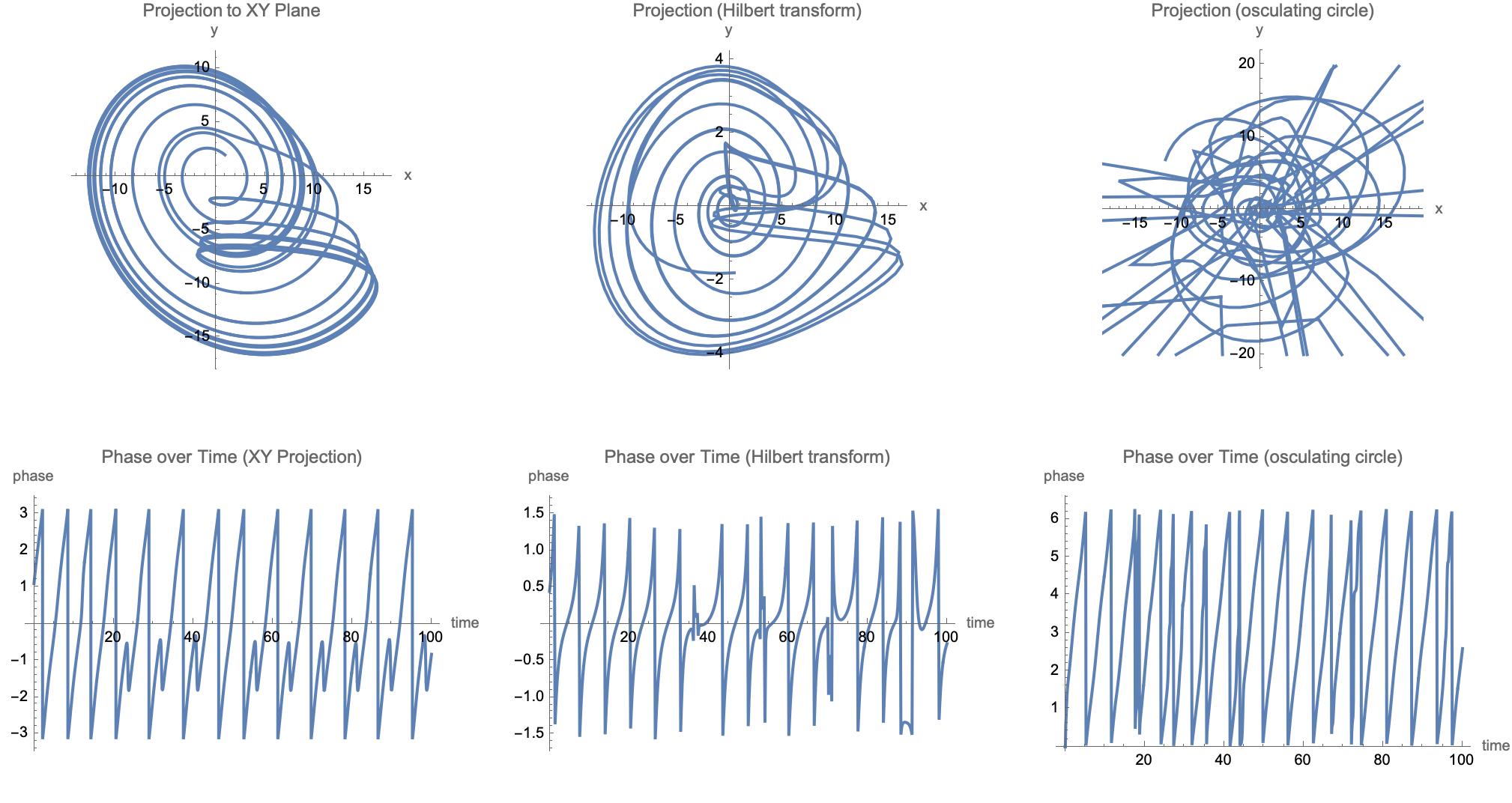}
    \caption{Comparison of different methods to determine the phase of the R\"ossler system. Top row: Planar trajectories reconstructed using (left) projection to the $xy$ plane, (middle) Hilbert transform of the $x$ component, and (right) osculating circle method. Bottom row: Corresponding phases over time using $\arctan2(y(t), x(t))$ for the first two methods and integration of $\omega(t)$ for the osculating circle method.}
    \label{fig:phasesbydefinitionnc}
\end{figure*}

Let us compare this to a non-phase-coherent case; we will change the parameters of the R\"ossler system to $a = 0.3$, $b = 0.4$, and $c = 7.5$. In that case, the projection method and the Hilbert based method do not produce a monotonously increasing phase. The method via the osculating circle, however, still gives a clear phase signal, Figure \ref{fig:phasesbydefinitionnc}.

As before, the 2 dimensional projections look cleaner for the standard methods, due to the large changes of the radius of the osculating circle, but the it is only our new method which provides a clear phase signal, i.e. only in the new method is the signal monotonously increasing, or in this case being reset by the modulo $2 \pi$ function.

\section{\label{sec:lorenz} Application to Lorenz Systems}

In this section, we apply the osculating circle method to the Lorenz system and a non-phase coherent Lorenz system driven by a Rössler system. These examples demonstrate the versatility and robustness of our method in handling different chaotic systems, which do not have a clear rotational centre.

\subsection{Lorenz System}

\begin{figure*}[htbp]
\centering
\includegraphics[width=\textwidth]{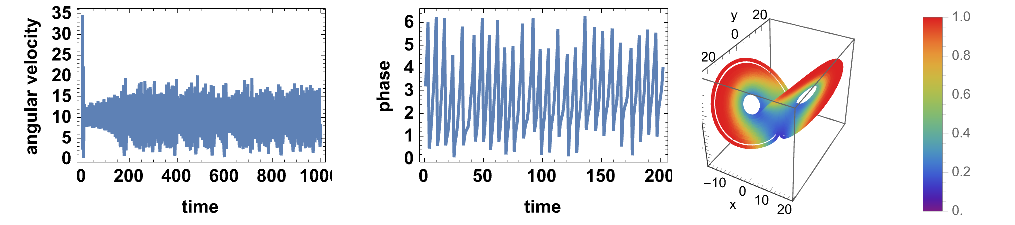}
\caption{Phase and angular speed representations for the Lorenz system. (left) Angular velocity, (middle) Instantaneous phases mod $2\pi$, (right) Attractor coloured according to phase speed.}
\label{fig:lorenz-results}
\end{figure*}

The Lorenz system is another classic model in chaotic dynamics, defined by the following differential equations:

\begin{align}
\frac{dx}{dt} &= \sigma (y - x), \notag \\
\frac{dy}{dt} &= x (r - z) - y, \label{lorenzeqns} \\
\frac{dz}{dt} &= xy - \beta z, \notag
\end{align}

where the standard parameters used are $\sigma = 10$, $r = 28$, and $\beta = 8/3$. These parameters yield a chaotic attractor, often referred to as the Lorenz butterfly, Figure \ref{fig:lorenz-results} (right).

Using the osculating circle method, we compute the curvature $\kappa(t)$ and the angular speed $\omega(t)$, which allows us to derive the instantaneous phase $\phi(t)$. This method provides a clear phase signal, without the need of any additional coordinate system transformations.

The Lorenz system does not orbit around a single point but rather has two lobes. Therefore, a direct phase definition via projection is not possible without additional transformations. Using the osculating circle, we obtain the phase directly without the need to determine a specific projection.

Figure \ref{fig:lorenz-results} shows the phase analysis for the Lorenz system. The left panel  presents the angular velocity, the middle panel shows the instantaneous phases mod $2\pi$, and the right panel displays the attractor coloured according to phase speed. The phase obtained via this method is a monotonously increasing scalar, providing a robust phase signal for synchronisation analysis.

\subsection{Lorenz System Driven by a Rössler System}

Next, we compute the phases of a non-phase coherent Lorenz system driven by a Rössler system. The respective system of differential equations is:

\begin{align}
\frac{dx_1}{dt} &= b + x_1(t) \left( x_2(t) - c \right), \notag \\
\frac{dx_2}{dt} &= -x_1(t) - x_3(t), \notag \\
\frac{dx_3}{dt} &= x_2(t) + a x_3(t), \notag \\
\frac{dy_1}{dt} &= -\sigma \left( y_1(t) - y_2(t) \right), \notag \\
\frac{dy_2}{dt} &= r \left( x_1(t) + x_2(t) + x_3(t) \right) - y_2(t) \notag \\
&\quad - \left( x_1(t) + x_2(t) + x_3(t) \right) y_3(t), \label{lorenzrossler} \\
\frac{dy_3}{dt} &= \left( x_1(t) + x_2(t) + x_3(t) \right) y_2(t) - \beta y_3(t), \notag
\end{align}

with parameters $a= 0.45$, $b= 2$, $c= 4$, $\sigma= 10$, $r = 28$, and $\beta= 8/3$.

\begin{figure}[htbp]
\centering
\includegraphics[width=0.35\textwidth]{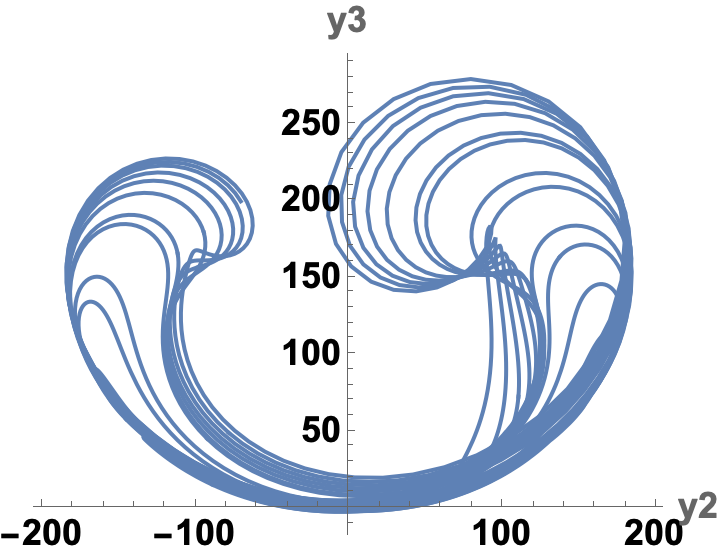}
\caption{Two-dimensional projection of the driven Lorenz system with parameters $a= 0.45$, $b= 2$, $c= 4$, $\sigma= 10$, $r = 28$, and $\beta= 8/3$.}
\label{fig:coupled-lorenz-2D}
\end{figure}

Using the osculating circle method, we derive the angular velocity and instantaneous phase for this coupled system.

\begin{figure*}[ht]
\centering
\includegraphics[width=\textwidth]{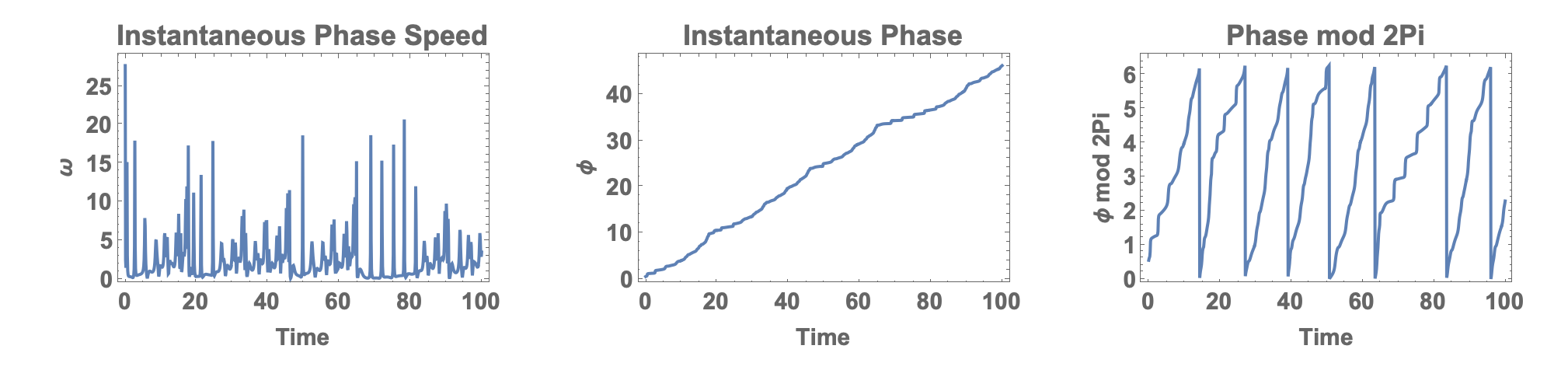}
\caption{Phase analysis of the Lorenz system driven by a Rössler system using the osculating circle method. (left) Instantaneous phase speed, (middle) Integrated phase, (right) Phase modulo $2\pi$.}
\label{fig:phaseforlorenzbyroessler}
\end{figure*}

Figure \ref{fig:coupled-lorenz-2D} displays the 2D projection of the driven Lorenz system. The attractor clearly shows that there is no single center about which the trajectory rotates. 
Using common approaches such as projection, it would be difficult to define a phase due to the lack of a rotation center. In the past, recurrence-based methods have been used to check for synchronisation in such cases, but a direct phase definition is problematic. 

Using the osculating circle method, however, the definition yields a clean and well-defined phase signal directly. Figure \ref{fig:phaseforlorenzbyroessler} presents the results of the phase analysis. The left panel shows the instantaneous phase speed, the middle panel  the integrated phase, and the right panel the phase modulo $2\pi$. These results highlight the method's robustness in providing a clear phase signal even in non-phase coherent systems.

This section illustrates the applicability of the osculating circle method in different chaotic systems, further demonstrating its versatility and robustness in phase analysis.

\section{\label{sec:sync} Application to Synchronisation Analysis}

Phase synchronisation in chaotic systems involves the process of adjusting phases of distinct systems into a coherent temporal structure. It is a phenomenon where the phases of interacting oscillators become locked, while their amplitudes may remain uncorrelated and chaotic. This concept is crucial for understanding complex interactions in various natural and technological systems, such as in laser dynamics, electronic circuits, chemical reactions, and biological rhythms \cite{Pikovsky2001}.

Phase synchronisation can be mathematically described as the entrainment of the phases of chaotic oscillators. Given two coupled oscillators with phases $\phi_1(t)$ and $\phi_2(t)$, phase synchronisation is achieved when the phase difference $\phi_1(t) - \phi_2(t)$ remains bounded for all time, despite the individual phases $\phi_1(t)$ and $\phi_2(t)$ growing without bound. This can be formally expressed as:
\[
|\phi_1(t) - \phi_2(t)| < \text{constant}.
\]
This differs from complete synchronisation, where not only the phases but also the amplitudes of the oscillators are synchronised, leading to identical states for both systems.

To illustrate the effectiveness of our approach, we examine two coupled R\"ossler systems:
\begin{align}
\frac{dx_1}{dt} &= -\omega y_1 - z_1 + \epsilon (x_2 - x_1),\notag \\
\frac{dy_1}{dt} &= \omega x_1 + a y_1,\notag  \\
\frac{dz_1}{dt} &= b + z_1 (x_1 - c), \label{coupledroesslers} \\
\frac{dx_2}{dt} &= -(\omega+\Delta) y_2 - z_2 + \epsilon (x_1 - x_2), \notag \\
\frac{dy_2}{dt} &= (\omega+\Delta) x_2 + a y_2, \notag  \\
\frac{dz_2}{dt} &= b + z_2 (x_2 - c) \notag
\end{align}

The parameters used in the simulation are: $a = 0.165, b = 0.2, c = 10; \omega = 0.97$. The parameter $\Delta$ introduces a frequency mismatch, and $\epsilon$ represents the coupling strength. Phase synchronisation in this context can be visualised using the concept of the {\em Arnold's tongue}, which shows regions in parameter space where synchronisation occurs.

Note that phase synchronisation is the process of adaptation of the phases of the systems. Both systems generally have different parameters and slightly different eigenfrequencies. When synchronisation takes place, the phases adjust, and in spite of different eigenfrequencies the systems lock their phases. This locking process involves the "slower" system (in terms of phase speed) accelerating, while the "faster" system decelerates. If two uncoupled systems have the same eigenfrequencies, their phases might appear to be locked, but there is no adaptation.

Using the osculating circle method, we observe the adaptation process through the excursions in the angular speed curve occurring at similar times for both systems. This method highlights how the angular speeds of the two systems adjust during synchronisation.

In the visualisation of Arnold's tongue, traditional methods use colours to represent phase differences. Small phase differences correspond to synchronisation, and, when a temperature colour map is used, low values (blue) indicate synchronisation. For the osculating circle method, we use the correlation between angular speeds $\omega$ to check for synchronisation. Higher values indicate synchronisation, so the colours are inverted compared to the traditional method, with high values indicating synchronisation.

\begin{figure*}[htbp]
\centering
\includegraphics[width=0.7\textwidth]{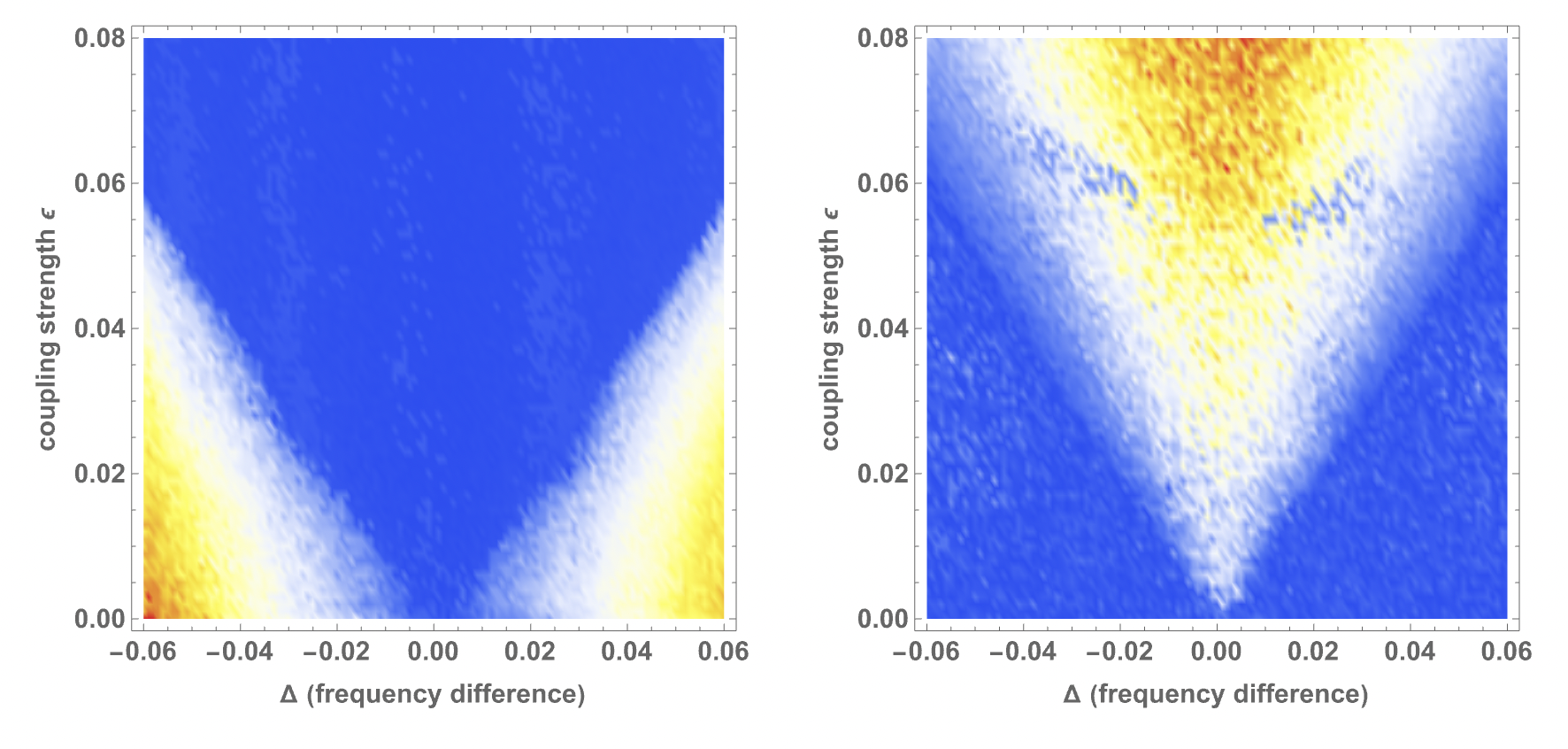}
\caption{Arnold’s tongue showing the resulting frequency difference (colour map) between two coupled chaotic R\"ossler oscillators. The x-axis represents the frequency difference of the uncoupled oscillators, and the y-axis represents the coupling strength. The figure demonstrates how varying the coupling strength allows the oscillators to adjust their phases and achieve synchronisation. Left: Traditional phase definition - The temperature colour map indicates the resulting frequency difference between the oscillators. Low temperatures (e.g. blue) indicate synchronisation. Right: Osculating circle method indicating angular speed correlation. As correlation between angular speeds is used, higher values, i.e. warmer colours, indicate synchronisation.}
\label{fig:arnoldstongues}
\end{figure*}

Our analysis, depicted in Figure \ref{fig:arnoldstongues}, contrasts the traditional phase definition with the osculating circle method. While traditional methods focus on phase differences, our approach emphasises angular speed correlations. The figure calculated via the osculating circle method contains faint structures, at $\Delta \pm 0.025$ and $\epsilon = 0.06 \pm 0.0025$. These indicate a decrease in correlation of the angular speeds of the two systems. Further analysis reveals that this region has been observed earlier when Arnold's tongues were studied with Joint Cross Entropies \cite{Marwan2007}. Closer analysis shows that these regions correspond to so-called lag synchronisation.

This comparative analysis highlights the osculating circle method's ability to handle non-linear dynamics and offer insights into transitional behaviours within chaotic systems. The results underscore the potential of this method in broadening our understanding of phase synchronisation in chaotic systems, extending its application across further scientific fields, when traditional phase definitions do not yield a phase signal.

\section{Higher-Dimensional Phase and Synchronisation Analysis}\label{sec:higherDims}

In this section, we explore the application of the osculating circle method to phase analysis and synchronisation in higher-dimensional coupled R\"ossler systems, extending the discussion from Section \ref{sec:highdim}. We focus on two coupled R\"ossler systems (Eqns \ref{coupledroesslers}), examined both as individual entities and as a unified six-dimensional system. This analysis emphasises the intricate interplay and synchronisation induced by coupling.

\begin{figure*}[htbp]
 \centering
  \includegraphics[width=0.6\textwidth]{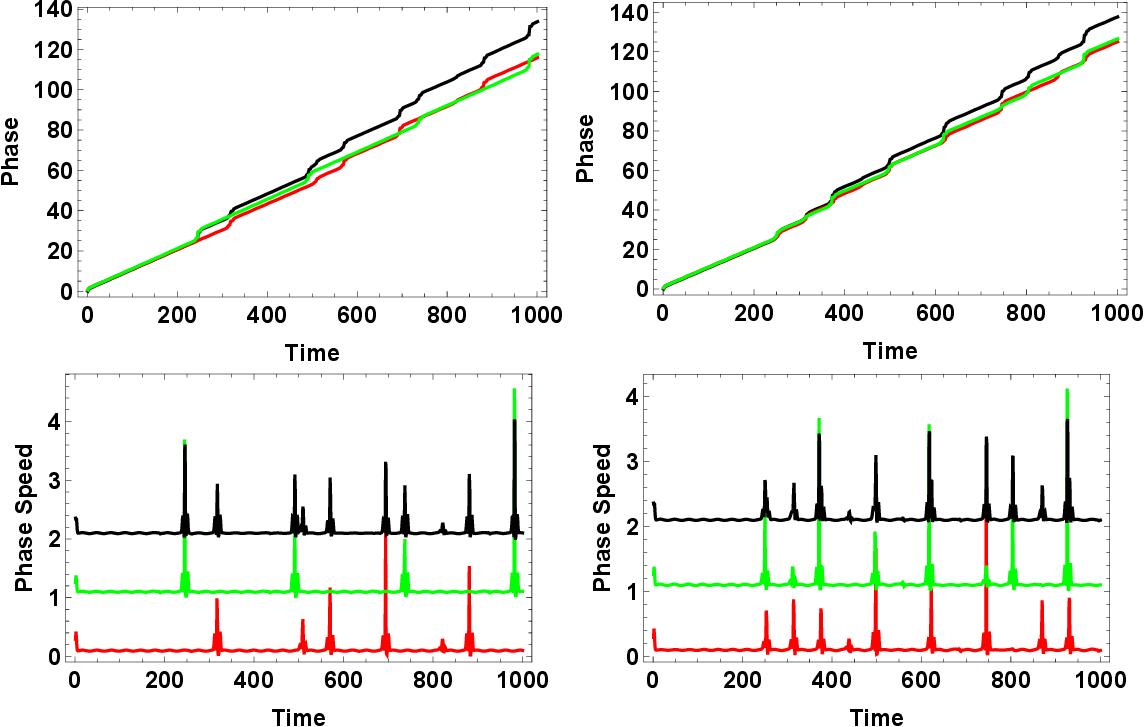}
  \caption{Depiction of phases and angular velocities for two coupled R\"ossler systems. (Left) non-synchronised systems; (Right) synchronised systems. Individual systems (System 1 in red, System 2 in green) and the aggregate system (in black) are shown. The upper panel traces the phase evolution, underscoring the synchronous phase adjustments in the combined system, mirroring abrupt changes in individual subsystems. The lower panel, displaying angular speeds with interpretive offsets, reveals alignment of high-curvature regions in the synchronised state, contrasting with the disjointed pattern in the non-synchronised state. (An offset of 1 was added to the second subsystem and an offset of 2 to the full system.)}
    \label{fig:phasestotalvspart}
\end{figure*}

Figure \ref{fig:phasestotalvspart} contrasts the phase behaviours of the two systems under different synchronisation states. In the non-synchronised state, with parameters set to $a = 0.165, b = 0.2, c = 10, \omega = 0.97, \Delta = 0.03, \epsilon = 0.005$, the individual subsystems (System 1 in red, System 2 in green) and the total system (in black) exhibit distinct phase patterns. The synchronised state, with parameters $a = 0.165, b = 0.2, c = 10, \omega = 0.97, \Delta = 0.03, \epsilon = 0.1$, demonstrates a notable alignment in phase increments, reflecting synchronised dynamics. The angular speed spikes in the lower panel, indicative of high-curvature areas, align in the synchronised state, contrasting the asynchronous pattern in the non-synchronised state.

This analysis reveals the capability of the osculating circle method in identifying and characterising synchronisation in complex, higher-dimensional systems. The phase growth of the total system mirrors rapid increases in either subsystem, and the alignment of high-curvature regions during synchronisation provides insights into the dynamical properties of these coupled systems. 

Moreover, this approach allows us to examine when two coupled systems effectively become a single unified system. By comparing the phases and angular speeds of the subsystems and the total system, we can gain a deeper understanding of the transition to synchronisation.

These results underscore the effectiveness of the osculating circle method for higher-dimensional phase analysis, making it a valuable tool for theoretical exploration and practical application in studying the dynamics of coupled systems.

\section{Implications and Future Directions}

The osculating circle method, initially formulated for chaotic systems, shows versatility and potential applicability across a range of dynamical systems. Its extension to non-autonomous systems can aid in understanding the behaviour of systems influenced by time-dependent external factors. This adaptation can provide new insights in control and synchronisation studies, facilitating system response analysis under variable conditions.
A promising application of the osculating circle method lies in data-driven models, and particularly in applications to time series analysis. The geometric perspective offered by this method can enhance interpretability, contributing to a better understanding of complex system dynamics. This can broaden the method’s applicability and support interdisciplinary research, integrating theoretical mathematics, physics, and applied computational science. The osculating circle method offers a new approach in phase analysis for chaotic systems, which can be extended to various phenomena like quantum chaos, fluid dynamics, and biological systems. Future research could explore its utility in these areas, potentially providing new insights into their chaotic behaviour.
This method’s geometric approach can refine existing theoretical frameworks, improving our understanding of chaos and phase synchronisation. It has interdisciplinary potential, particularly in areas such as neuroscience, climate science, and engineering, where understanding and controlling chaotic systems is crucial \cite{Lakshmanan2003}. In neuroscience, it could help unravel complexities in neural dynamics and brain connectivity. In engineering, it suggests more advanced control systems for chaotic processes in mechanical and electronic systems. 

\section{\label{sec:level4}Conclusion}

This study presents a new tool for the analysis of oscillatory systems, including chaotic systems. The osculating circle method, based on differential geometry, provides a natural, geometrically defined phase for these systems. Unlike traditional methods, it does not rely on arbitrary transformations to establish phase, which is particularly beneficial in systems where phase coherence is less apparent or in higher-dimensional settings.
Our approach is applicable for both theoretical understanding and practical data analysis, offering a robust tool for examining the intricate dynamics of various complex systems. It is particularly effective in handling non-phase-coherent and multi-dimensional systems, thus broadening the range of dynamical systems that can be effectively analysed.
In summary, the osculating circle method represents a technique to analyse and interpret oscillatory systems. It has the potential to enhance understanding of these systems, both in theory and through empirical data analysis, opening new avenues for research and application across scientific disciplines.

\bibliographystyle{plain}

\bibliography{OsculatingCircle.bib}

\end{document}